**Title**

Crystal structures of LaO$_{1-x}$F$_x$BiS$_2$ ($x \sim 0.23, 0.46$): effect of F doping on distortion of Bi–S plane

**Authors**

Akira Miura*, Masanori Nagao, Takahiro Takei, Satoshi Watauchi, Isao Tanaka, Nobuhiro Kumada

*Corresponding Author

E-mail address: amiura@yamanashi.ac.jp

Postal address: University of Yamanashi, Center for Crystal Science and Technology

Miyamae 7-32, Kofu 400-8511, Japan

Telephone number: (+81)55-220-8614

Fax number: (+81)55-254-3035

**Abstract**

The crystal structures of superconducting $LaO_{1-x}F_xBiS_2$ ($x \sim 0.23, 0.46$) were determined by single-crystal X-ray diffraction analysis. Their space group was $P4/nmm$. Distortion of the Bi–S plane changed when the F content was increased from 0.23 to 0.46, and a nearly flat Bi–S plane was formed at $x \sim 0.46$. Computational calculations supported this effect of F doping on distortion of Bi–S plane. $LaO_{1-x}F_xBiS_2$ with higher F contents were computationally predicted to be thermodynamically more unstable under ambient pressure. We discussed the bonding, conductivities, and synthetic routes of $LaO_{1-x}F_xBiS_2$.

# 1. Introduction

The bonds between metals and anions determine crystal structures and affect physical properties, and recent advances in mixed-anion chemistry have been based on understanding and controlling the effects of anions [1-3]. $LnO_{1-x}F_xBiS_2$ (Ln: La, Ce, Pr, Nd, or Yb) are superconducting mixed-anion layered compounds [4-13]. Among these, $LaO_{1-x}F_xBiS_2$ is reported to have a slightly distorted Bi–S plane, and shows superconducting transitions up to ~10 K [4-8]. Distortion of the Bi–S layers could be attributed to a filled s-orbital of Bi; most of the coordination around heavy s- or p-block elements such as $Sn^{2+}$, $Pb^{2+}$, and $Bi^{3+}$ is highly asymmetric [14-16]. Substitution of O by F is considered to induce carrier into the Bi–S layer, and this is believed to trigger superconductivity [4-12]. The effect of F doping on the lattice parameters and the superconducting transition temperature has been reported for $LaO_{1-x}F_xBiS_2$ [6]. However, the effect of this substitution has been still unclear; crystallographic studies have mostly been performed on polycrystalline samples with estimated concentrations of F [4-7]. Rietveld analysis of powder X-ray powder diffraction by using CuKα radiation is difficult to give accurate position and displacement parameters especially for light elements (F, O, S). There has been one report on single-crystal X-ray diffraction analysis of F-substituted $NdO_{1-x}F_xBiS_2$ ($x$ ~ 0.39), but no structural investigation of different F contents has been performed [12]. Therefore, the effect of F dopant on distorted Bi–S plane is an open question, which could be the key for the appearance of superconductivity.

In this paper, we report nearly flat and slightly zigzag Bi–S planes in $LaO_{1-x}F_xBiS_2$ with different amounts of F dopant, based on single-crystal X-ray diffraction analysis and first-principle calculations. We show how different degrees of substitution of O by F affect the crystal structures and thermodynamic stabilities of $LaO_{1-x}F_xBiS_2$, and discuss their bonding, superconductivities, and synthetic routes.

## 2. Experimental and computational details

The synthesis, chemical analysis, and transport properties of single crystals of $LaO_{1-x}F_xBiS_2$ were described in our previous report [8]. Briefly, plate-shaped metallic crystals were synthesized using a CsCl/KCl flux at 873–1073 K in an evacuated quartz tube. $La_2S_3$, Bi, $Bi_2O_3$, $Bi_2S_3$, and $BiF_3$ powders mixed in nominal compositions of $LaO_{0.7}F_{0.3}BiS_2$ and $LaO_{0.3}F_{0.7}BiS_2$ were used as starting materials. The F contents of two types of single crystal were determined to be $x \sim 0.23$ and 0.46 by electron probe microanalysis (EPMA).

Structural analysis of single crystals was performed according to the previous report on F-doped NdOBiS$_2$ single crystal [12]. X-ray diffraction was performed using a RigakuXTALAB-MINI diffractometer with graphite-monochromated Mo Kα radiation. The data were corrected for Lorentz and polarization effects. The crystal structures were solved and refined using computer programs from the Crystal Structure crystallographic software package with SHELXL-97 [17,18]. The structures were drawn using VESTA [19].

First-principle calculations were performed based on the plane-wave pseudopotential strategy, using the Vienna *Ab initio* Simulation Package (VASP) [20]. The projector-augmented wave approach [21,22] and the generalized PBE-type gradient approximation were used [23]. We chose an energy cutoff of 500 eV. The $\sqrt{2} \times \sqrt{2} \times 1$ supercells of LaO$_{1-x}$F$_x$BiS$_2$ ($x$ = 0, 0.25, 0.50, 0.75, and 1.00) were constructed with the space group $P$1 by using previously reported structure [4] and used as initial structures. We employed one model for each F contents of LaO$_{1-x}$F$_x$BiS$_2$ with ordered occupancies because it is hard to perform the calculation assuming random occupancies of O/F. The $k$-point 14 × 14 × 6 grids were automatically generated [24], and all the atomic positions, volume, and shape were optimized. The structures of Bi, Bi$_2$O$_3$, Bi$_2$S$_3$, BiF$_3$, La$_2$O$_3$, La$_2$S$_3$, and LaF$_3$ were also optimized by the similar way using reported unit cells as initial structures with at least 54 $k$ points.

## 3. Results and discussion

The systematic absences in LaO$_{1-x}$F$_x$BiS$_2$ (x ~ 0.23, 0.46) shown by the XRD patterns suggested that the space group was *P*4/*nmm*. Structural refinement was performed using the structural data for NdOBiS$_2$, which has the space group *P*4/*nmm*, as the initial model [12]. Repeated least-squares refinements determined the positions and anisotropic atomic displacement parameters. Among the several crystals, the refinements suggested little or small amounts of Bi vacancy (0-3 %) but no vacancies in other sites. This result agrees with the compositions determined by EPMA analysis within the error (La:Bi:S = 1.01±0.05:0.99±0.05:2.00) [8]. The F content followed the ratio determined by EPMA (*x* ~ 0.23, 0.46) [8] since this could not be determined using X-ray analysis: there is little difference between the scattering factors of O and F. We therefore report the refinements with fixed occupancies of La, Bi, O/F, and S as unity though we cannot deny the possibility of small amounts of Bi vacancy. The final *R*1 values for LaO$_{1-x}$F$_x$BiS$_2$ at *x* ~ 0.23 and 0.46 are 0.0360 and 0.0354, respectively. Details of the data collection and refinement are summarized in Tables 1 and 2. The bond valence sum (BVS) of O/F site at *x* ~ 0.46 is smaller than that at *x* ~ 0.23, which supports more incorporation of F into O site. The atomic displacement parameters of S(1) are relatively large for $x \sim 0.23$ and 0.46; this is also found for $NdO_{1-x}F_xBiS_2$ ($x \sim 0.39$) [12].

Figure 1 shows the lattice parameters of $LaO_{1-x}F_xBiS_2$ single crystals. These derived from polycrystalline samples with/without high-pressure annealing [4-7] and the values predicted using first-principles calculations are also shown for comparison. Single-crystal X-ray diffraction analysis shows that increasing the F content from 0.23 to 0.46 slightly increases the *a*-axis lattice parameter [$x \sim 0.23$: $a = 4.057(3)$ Å; $x \sim 0.46$: $a = 4.063(2)$ Å] and shrinks that of the *c*-axis [$x \sim 0.23$: $c = 13.547(4)$ Å; $x \sim 0.46$: $c = 13.345(4)$ Å]. These lattice parameters agree with the data for the polycrystalline samples. The experimental lattice parameters determined from single-crystal and polycrystalline samples are similar to those predicted by computational calculations up to $x \sim 0.5$. However, with increasing F content from 0.5, the computationally predicted lattice parameters show significant differences from the experimental values. Only the *c*-axis of the polycrystalline sample with high-pressure annealing follows the computational trend. The growth of single crystals at $x > 0.5$ was unsuccessful under ambient pressure, even when the crystals were synthesized from starting materials with high F contents ($x = 0.7$ and 0.9) [8]. While polycrystalline samples with high-pressure annealing shows superconductivity in the wide range (0.2-0.7) [4,6], single-crystal and polycrystalline samples without high-pressure annealing are reported as superconductors in the narrow range near $x = 0.5$ [4-8].

Figure 2 shows the crystal structure of $LaO_{1-x}F_xBiS_2$. This layered structure can be seen as a stack of distorted Bi–S planes, S planes, and edge-sharing $La_4$(O/F) tetrahedral slabs. Six-coordinated Bi atoms form three kinds of Bi–S bonds. One is along the *ab* plane, and the other two are along the *c*-axis.

Table 3 shows a summary of selected bond distances and angles determined by single-crystal X-ray diffraction analysis. The interatomic distances of in-plane Bi–S(1) at $x \sim 0.23$ and 0.46 are comparable [$x \sim 0.23$: 2.870(2) Å, $x \sim 0.46$: 2.8730(14) Å]. This is slightly longer than the in-plain Bi–S(1) distance in $NdO_{1-x}F_xBiS_2$ [$x \sim 0.39$: 2.827(2) Å]. The interplanar Bi–S(1) bond is much longer, and the Bi–S(2) bond is shorter than that of the in-plane Bi–S(1) bond in $LaO_{1-x}F_xBiS_2$ ($x \sim 0.23$ and 0.46). Such large deviations in bond lengths are found in $Bi_2S_3$ with a layered structure [25]. The in-plane S–Bi–S angle at $x \sim 0.23$ deviates slightly from 180° [177.2(4)°], whereas that at $x \sim 0.46$ is nearly flat [180.8(3)°], indicating less distortion of the Bi–S plane. Doping with F therefore not only increases the carrier concentration but also changes the distortion of the Bi–S plane. The mean bond lengths in the Bi–S(1) plane predicted by first-principles calculations ($x = 0.25$: 2.878 Å; $x = 0.5$: 2.885 Å) were close to the experimental values, and an almost flat Bi–S plane was also predicted at $x = 0.5$; the S(1)–Bi–S(1) angles were 173.4° and 179.3° at $x = 0.25$ and 0.5, respectively. However, unusual deviations were found in the interlayer distances of Bi–S(1) ($x = 0.25$: 3.223 Å; $x = 0.5$: 3.091 Å) and Bi–S(2) ($x = 0.25$: 2.591Å; $x = 0.5$: 2.658 Å) at $x = 0.5$; similar results have been reported in the other computational study [27]. The bond lengths of the in-plane Bi–S(1) of polycrystalline LaO$_{1-x}$F$_x$BiS$_2$ ($x \sim 0.5$; Bi–S(1): 2.870 Å [4], 2.885 Å [5]) determined by Rietveld analysis are comparable to those obtained from the single-crystal data, but their in-plane angles are different (S(1)–Bi–S(1): 186.0° [4], 188.5° [5]). This may be attributed to the high-pressure annealing [4] and/or obscure F contents in the powder sample, and further examinations are necessary. The interatomic La–O/F lengths in LaO$_{1-x}$F$_x$BiS$_2$ ($x \sim 0.23$ and 0.46) are 2.4117(16) and 2.4338(18) Å, respectively. These are longer than that in NdO$_{1-x}$F$_x$BiS$_2$ [$x \sim 0.39$: 2.3704(17) Å], because the ionic radius of La$^{3+}$ (1.160 Å [26]) is larger than that of Nd$^{3+}$ (1.109 Å [26]). The La-O/F lengths in LaO$_{1-x}$F$_x$BiS$_2$ are slightly longer than those in 5 atom% F-doped LaOFeAs (2.3668(3) Å [1]).

Figure 3 shows the temperature dependence of resistivity of LaO$_{1-x}$F$_x$BiS$_2$ ($x \sim 0.23$ and 0.46). While LaO$_{1-x}$F$_x$BiS$_2$ ($x \sim 0.23$) shows semiconducting behavior in normal state region, LaO$_{1-x}$F$_x$BiS$_2$ ($x \sim 0.46$) exhibits metallic one. This suggests that the increased amount of substitution amount of O by F enhanced the carrier concentration, and semiconductor-metal transition likely occurs in the F amount between 0.23 and 0.46. Although $LaO_{1-x}F_xBiS_2$ ($x \sim 0.46$) shows zero resistivity below ~3 K, $LaO_{1-x}F_xBiS_2$ ($x\sim0.23$) does not show it down to 2 K.

Figure 4 shows the computationally predicted thermodynamic stabilities of $LaO_{1-x}F_xBiS_2$ with different F contents. We investigated two decomposition reactions, according to the synthetic routes [4,8].

$LaO_{1-x}F_xBiS_2 \rightarrow 1/2La_2S_3 + 1/6Bi_2S_3 + x/3BiF_3 + x/3Bi + (1-x)/3Bi_2O_3$ (1)

$LaO_{1-x}F_xBiS_2 \rightarrow (1-x)/3La_2O_3 + (1+x)/6La_2S_3 + x/3LaF_3 + x/3Bi + (3-x)/6Bi_2S_3$ (2)

An increase in the F content is found to decrease the thermodynamic stability of $LaO_{1-x}F_xBiS_2$. The decomposition reaction (1) is endothermic for all F contents, and therefore is unlikely to occur. However, reaction (2) is exothermic when the F content exceeds ~0.5. Thus, it is difficult to synthesize $LaO_{1-x}F_xBiS_2$ with high F contents by a high-temperature method under ambient pressure.

The increase in the F content changes the distortion of the Bi–S plane, and an almost flat Bi–S plane is formed at $x \sim 0.46$. Theoretical study supports the distortion of the Bi–S plane. Zero resistivity is only observed for $LaO_{1-x}F_xBiS_2$ of $x \sim 0.46$ having nearly flat Bi–S plane. Although it is not clear how superconductivity is induced in LaO$_{1-x}$F$_x$BiS$_2$ materials, structural changes compete with superconductivity. Since the in-plane Bi–S distances are comparable for $x \sim 0.23$ and $0.46$, a nearly flat Bi–S plane should enhance hybridization of the Bi $6p_x/6p_y$ and S $3p_x/3p_y$ orbitals. Theoretical investigations shows that the Bi $6p_x/6p_y$ and S $3p_x/3p_y$ orbitals form quasi-one-dimensional bands intersecting the Fermi level [27,28]. Thus, Fermi surface would be changed by Bi–S bond. Single crystal X-ray diffraction analysis shows relatively large displacement parameters of S(1) regardless of F content, and in-plane displacement of S(1) is predicted to be related to charge-density-wave instability [27]. Therefore, flat Bi–S plane with relatively large displacement parameters of S(1) determined by single-crystal X-ray diffraction analysis can be an important factor for its superconductivity. The thermodynamic stability of LaO$_{1-x}$F$_x$BiS$_2$ predicted by first-principles calculations shows that the solubility limit of F is ~0.5 under ambient pressure. This explains the experimental finding that impurity phases above ~0.5 are formed in high-temperature synthesis [4], and the synthesis of single crystals with high F contents is unsuccessful under vacuum from starting materials with high F contents [8]. Growth of LaO$_{1-x}$F$_x$BiS$_2$ single crystals under high pressure may enable further investigation of LaO$_{1-x}$F$_x$BiS$_2$ compounds, especially in highly F-doped regions.

## 4. Conclusion

The crystal structures of LaO$_{1-x}$F$_x$BiS$_2$ ($x \sim$ 0.23 and 0.46) were determined by single-crystal X-ray diffraction analysis. Slightly zigzag and nearly flat Bi–S planes were found in LaO$_{1-x}$F$_x$BiS$_2$ at $x \sim$ 0.23 and 0.46, respectively. The effect of the substitution amount of O by F on the lattice parameters and in-plane Bi–S angle was supported by first-principle calculations. Therefore, doping of F not only increases the carrier concentration but also changes the distortion of the Bi–S plane. A flat Bi–S plane would result in better hybridization of the $p_x/p_y$ orbitals of Bi and S, and this can be related to the appearance of superconductivity. Thermodynamic predictions suggested that LaO$_{1-x}$F$_x$BiS$_2$ with higher contents of F above $x \sim$ 0.5 was not stable under ambient pressure. Further investigations of the crystal structures of various BiS$_2$-based materials will enable a deeper understanding of how mixed anions affect the bonding, crystal structure and properties.

**Supporting information**

CIF data, and additional figure and table for computational results are available.

**Figure Captions**

Figure 1. Relationship between F content ($x$) and lattice parameters of $LaO_{1-x}F_xBiS_2$. The figure shows the lattice parameters of single crystals prepared by high-temperature synthesis under vacuum (single crystal: this work), of polycrystalline samples prepared by high-temperature synthesis under vacuum (polycrystal) [4-7], and of polycrystalline samples prepared by high-temperature synthesis with subsequent high-pressure annealing [polycrystal (HP)] [4,6], and those predicted by first-principles calculations using the Vienna *Ab initio* Simulation Package (VASP) code (computational: this work). The $x$ contents of the single-crystal and polycrystalline samples were respectively determined by electron probe microanalysis and estimated from the compositions of the starting materials. The filled markers represent the appearance of superconductivity confirmed by zero resistivity at 2 K.

Figure 2. Crystal structure of $LaO_{1-x}F_xBiS_2$.

Figure 3. Temperature dependence of resistivity of $LaO_{1-x}F_xBiS_2$ ($x \sim 0.23, 0.46$) single crystals.

Figure 4. Relationship between F content, $x$, of $LaO_{1-x}F_xBiS_2$ and calculated energies of two decomposition reactions. Reaction 1: $LaO_{1-x}F_xBiS_2 \rightarrow 1/2La_2S_3 + 1/6Bi_2S_3 + x/3BiF_3 + x/3Bi + (1-x)/3Bi_2O_3$; Reaction 2: $LaO_{1-x}F_xBiS_2 \rightarrow (1-x)/3La_2O_3 + (1+x)/6La_2S_3 + x/3LaF_3 + x/3Bi + (3-x)/6Bi_2S_3$.

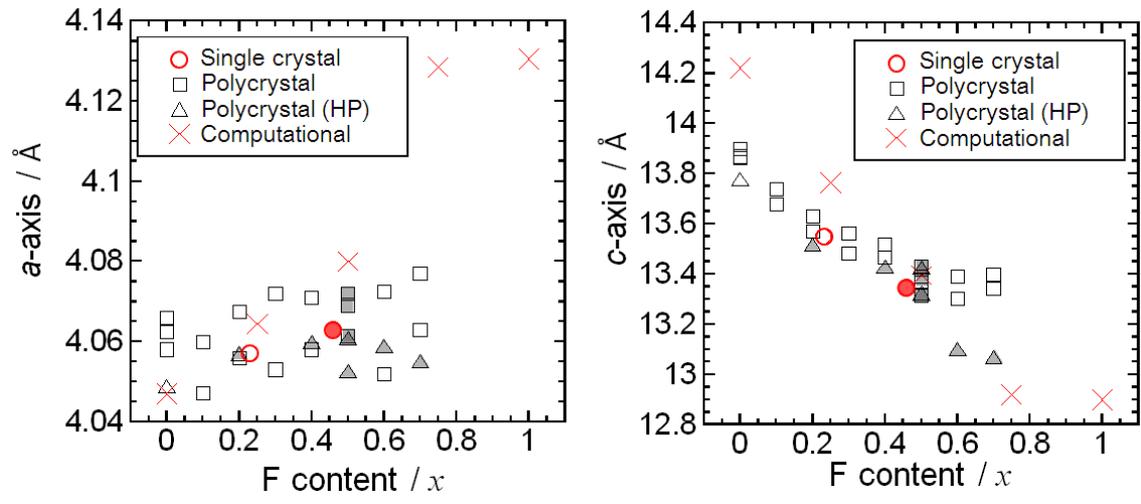

Figure 1

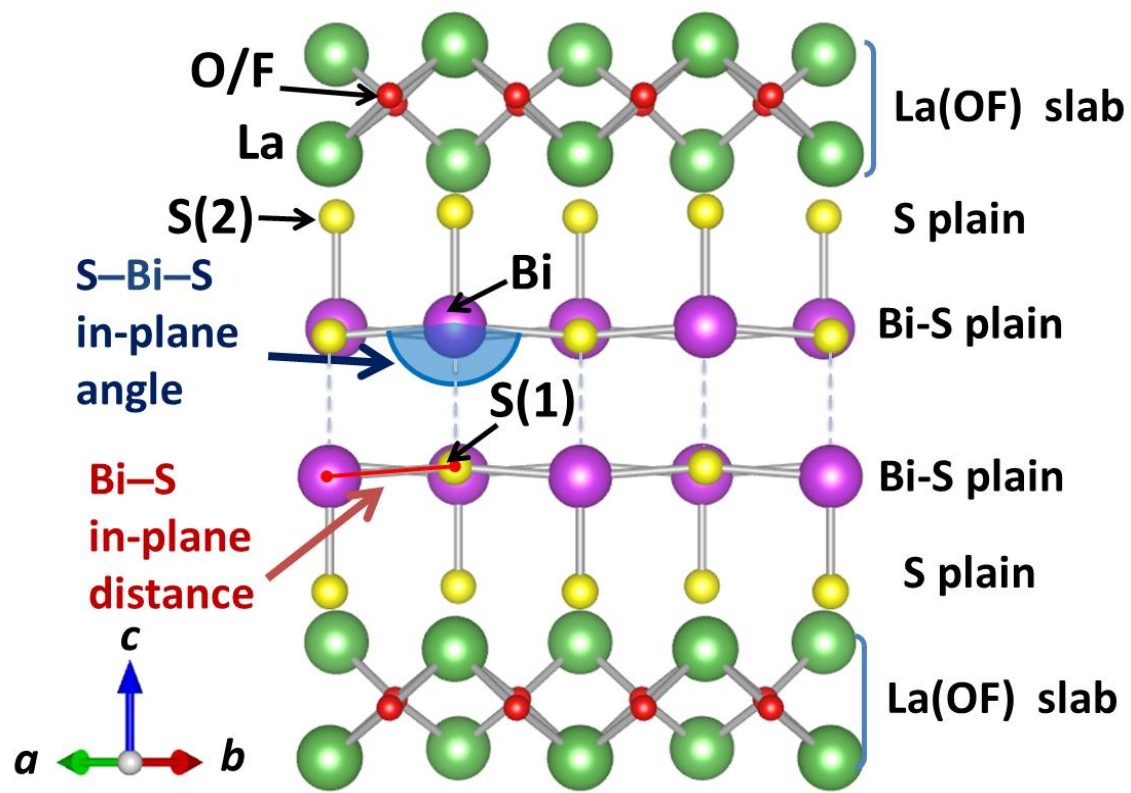

Figure 2

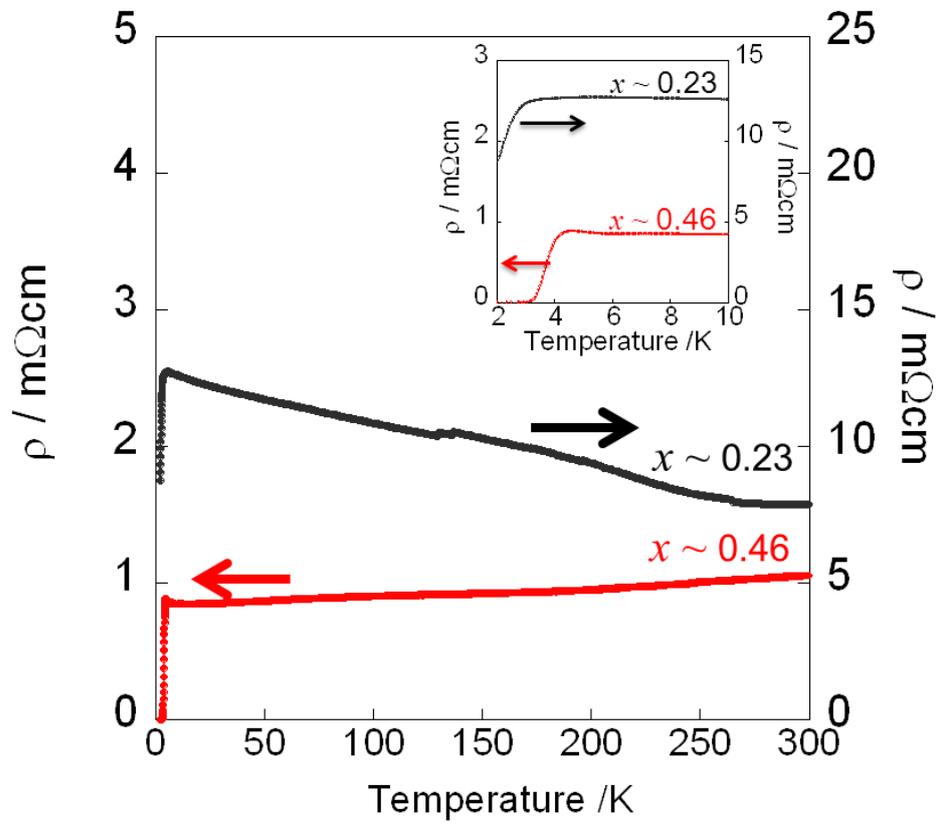

Figure 3

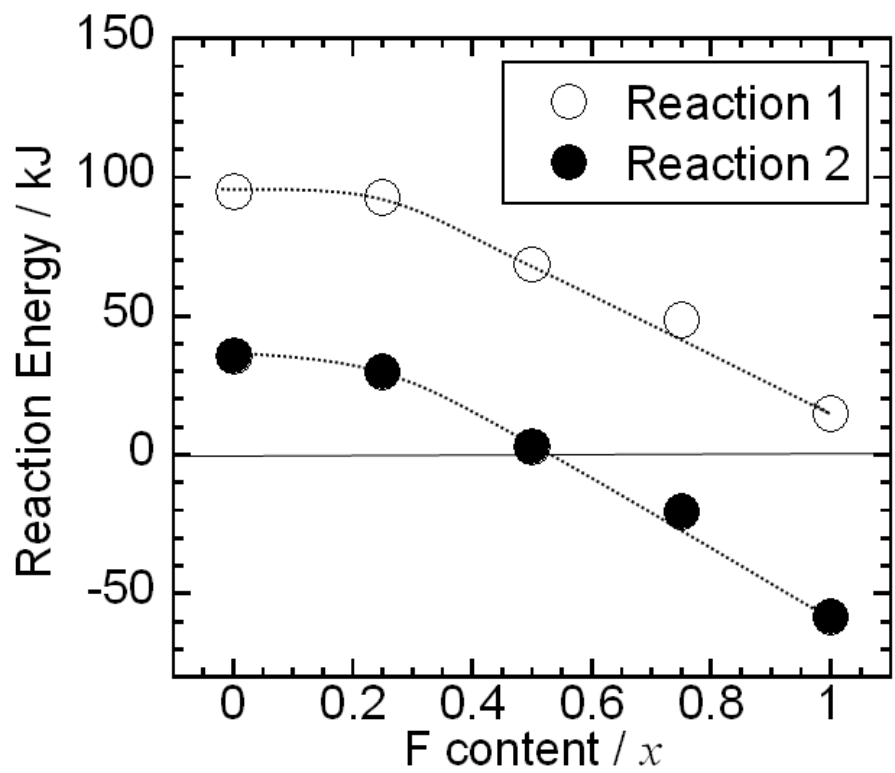

Figure 4

Table 1 Crystal data and intensity collections for LaO$_{1-x}$F$_x$BiS$_2$ single crystals ($x$~0.23, 0.46)

| Empirical Formula | LaO$_{1-x}$F$_x$BiS$_2$ ($x$~0.23) | LaO$_{1-x}$F$_x$BiS$_2$ ($x$~0.46) |
| --- | --- | --- |
| Crystal System | Tetragonal | Tetragonal |
| Lattice Parameters | $a$ = 4.057(3) Å | a = 4.063(2) Å |
| | $c$ = 13.547(4) Å | c = 13.345(4) Å |
| | $V$ = 223.0(3) Å$^3$ | V = 220.3(2) Å$^3$ |
| Space group | P4/$nmm$ (#129) | P4/$nmm$ (#129) |
| Z value | 2 | 2 |
| Diffractometer | Rigaku Mercury375R/M CCD (XtaLAB mini) | |
| Radiation | MoKα ($\lambda$=0.71075 Å) graphite monochromated | |
| Temperature | 24(3) °C | 24(3) °C |
| Scan type | $\chi$-2$\phi$ | |
| No. of total reflections | 2248 | 2257 |
| No. of unique reflections | 196 | 195 |
| No. of variables | 16 | 16 |
| Residuals: $R$1 ($I$>2.00σ($I$)) | 0.0360 | 0.0354 |
| Residuals: $R$ (All reflections) | 0.0397 | 0.0386 |
| Residuals: $wR$2 (All reflections) | 0.0877 | 0.0723 |
| Goodness of Fit Indicator | 1.07 | 1.04 |

Table 2 Positional, isotropic thermal parameters and bond valence sum (BVS) of LaO$_{1-x}$F$_x$BiS$_2$ ($x$~0.23, 0.46) determined by single-crystal X-ray diffraction analysis

| Atom | Site | $x$ | $y$ | $z$ | $B_{eq}$ | BVS |
|---|---|---|---|---|---|---|
| LaO$_{1-x}$F$_x$BiS$_2$ ($x$~0.23) | | | | | | |
| Bi | 2$c$ | 1/4 | 1/4 | 0.12549(8) | 1.21(5) | 3.06 |
| La | 2$c$ | -1/4 | -1/4 | 0.40371(14) | 0.97(6) | 3.06 |
| S(1) | 2$c$ | -1/4 | -1/4 | 0.1204(7) | 2.09(20) | 1.91 |
| S(2) | 2$c$ | 1/4 | 1/4 | 0.3113(5) | 0.87(13) | 2.23 |
| O/F | 2$b$ | -1/4 | -3/4 | 1/2 | 0.5(3) | 1.99 |
| LaO$_{1-x}$F$_x$BiS$_2$ ($x$~0.46) | | | | | | |
| Bi | 2$c$ | 1/4 | 1/4 | 0.12220(8) | 1.22(4) | 3.01 |
| La | 2$c$ | -1/4 | -1/4 | 0.39956(14) | 1.25(5) | 2.93 |
| S(1) | 2$c$ | -1/4 | -1/4 | 0.1235(6) | 2.2(2) | 1.91 |
| S(2) | 2$c$ | 1/4 | 1/4 | 0.3119(5) | 1.07(15) | 2.26 |
| O/F | 2$b$ | -1/4 | -3/4 | 1/2 | 0.6(4) | 1.77 |

*$B_{eq}$: Equivalent isotropic atomic displacement parameter

Table 3 Selected interatomic distances (Å) and angles (º) in $LaO_{1-x}F_xBiS_2$ determined by single-crystal X-ray diffraction analysis

|  | $LaO_{1-x}F_xBiS_2$ ($x$~0.23) | $LaO_{1-x}F_xBiS_2$ ($x$~0.46) |
| --- | --- | --- |
| Distance |  |  |
| Bi–S(1) [in-plane] × 4 | 2.870(2) | 2.8730(14) |
| Bi–S(1) [interplane] | 3.331(11) | 3.276(9) |
| Bi–S(2) | 2.514(7) | 2.534(6) |
| La–O/F ×4 | 2.4117(16) | 2.4338 (13) |
| La–S ×4 | 3.130(3) | 3.102(2) |
| Angle |  |  |
| S(1)–Bi–S(1) [in-plane] | 177.2(4) | 180.8(3) |
| S(1)–Bi–S(1) [interplane] | 89.967(9) | 89.997(3) |
| S(1)–Bi–S(2) | 91.38(19) | 89.59(16) |
| La–O/F–La | 114.51(9) | 113.17(8) |
| La–S–La | 80.79(9) | 81.82 (8) |